\documentclass[11pt]{article}%
\usepackage{amsfonts,bm}
\usepackage{mathrsfs}
\usepackage{amsmath}
\usepackage{graphicx}
\usepackage{hyperref}
\usepackage{verbatim}
\usepackage{color}
\usepackage{amsfonts}
\usepackage{amssymb}
\usepackage{xcolor}%
\providecommand{\U}[1]{\protect\rule{.1in}{.1in}}
\csname @addtoreset\endcsname{equation}{section}
\newcommand{\beq}{\begin{equation}}
\newcommand{\eeq}{\end{equation}}
\newcommand{\bea}{\begin{eqnarray}}
\newcommand{\eea}{\end{eqnarray}}
\newcommand{\ba}{\begin{array}}
\newcommand{\ea}{\end{array}}
\newcommand{\bit}{\begin{itemize}}
\newcommand{\eit}{\end{itemize}}

\newcommand{\complesso}{{\ \hbox{{\rm I}\kern-.6em\hbox{\bf C}}}}
\newcommand{\reale}{{\hbox{{\rm I}\kern-.2em\hbox{\rm R}}}}

\newcommand{\1}{ \,  \raisebox{+0.14em}{{\hbox{{\rm \scriptsize ]}} \raisebox{-0.2em}{\kern-.8em\hbox{1}}}} \, }

\begin{document}

\title{Analytic meronic black holes, gravitating solitons and higher spins in the
Einstein $SU(N)$-Yang-Mills theory}
\author{Fabrizio Canfora$^{1, 2}$, Andr\'es Gomberoff$^{1, 2}$, Marcela Lagos$^{3}$, Aldo
Vera$^{3}$\\$^{1}$\textit{Centro de Estudios Cient\'{\i}ficos (CECS), Casilla 1469,
Valdivia, Chile} \\
$^{2}$\textit{Facultad de Ingenier\'ia y Tecnolog\'ia, Universidad San Sebasti\'an, General Lagos 1163,}\\ \textit{Valdivia 5110693, Chile} \\
$^{3}$ \textit{Instituto de Ciencias F\'isicas y Matem\'aticas, Universidad
Austral de Chile,}\\\textit{ Casilla 567, Valdivia, Chile} \\{\small canfora@cecs.cl, gomberoff@cecs.cl, marcela.lagos@uach.cl,
aldo.vera@uach.cl}}
\maketitle

\begin{abstract}
We construct meronic black holes and solitons in the Einstein $SU(N)$%
-Yang-Mills theory in $D=4$ and $D=5$ dimensions. These analytical solutions
are found by combining the generalized hedgehog ansatz with the Euler
parameterization of the $SU(N)$ group from which the Yang-Mills equations are
automatically satisfied for all values of $N$ while the Einstein equations can
be solved analytically. We explicitly show the role that the color number $N$
plays in the black hole thermodynamics as well as in the gravitational spin
from isospin effect. Two remarkable results of our analysis are that, first,
meronic black holes can be distinguished by colored black holes by looking at
the spin from isospin effect (which is absent in the latter but present in the
former). Second, using the theory of non-embedded ansatz for $SU(N)$
together with the spin from isospin effect, one can build fields of arbitrary
high spin out of scalar fields charged under the gauge group. Hence, one can
analyze interacting higher spin fields in asymptotically flat space-times
without ``introducing by hand" higher spin fields. Our analysis also discloses
an interesting difference between the spin from isospin effect in $D=4$ and in
$D=5$.

\end{abstract}

\newpage

\tableofcontents

\section{Introduction}

Yang-Mills (YM) theory is one of the main ingredients of the
standard model which up to now has been phenomenologically extremely
successful. Since the main open problems in high energy physics such as color
confinement are non-perturbative in nature, it is of great interest to analyze
topologically non-trivial configurations of the YM theory which are believed
to play a fundamental role in the non-perturbative phase of the theory (see
\cite{greensite,DualSC,manton,BaMa,shifman1,shifman2,weinberg,NP,NP1,NP2,NP3,NP3.1,NP4,NP5} and references therein).

A very interesting class of configurations that play an important role in the
non-perturbative phase of the YM theory are the so-called
merons\footnote{Although the name ``meron" is generally used to describe
Euclidean solutions, in this work we will call merons to configurations with
$\lambda= 1/2 $ in Lorentzian space-time, which we will show in the following
sections.} introduced in \cite{merons1}. One of the characteristics of
merons is that they can always be brought in the form $A=\lambda\widetilde{A}%
$, where $\widetilde{A}$ is a pure gauge field. Since such an ansatz would be
trivial in Abelian gauge theories, merons are genuine non-Abelian
configurations. It is known that merons connect different topological sectors
of the theory and these are related to instantons \cite{meronsconfinement1,meronsconfinement2,meronsconfinement3,meronsconfinement4}. Also, lattice studies show that, as far as
confinement is concerned, merons play a very important role, as can be seen in
\cite{meronsconfinement1,meronsconfinement2,meronsconfinement3}. The existence of merons can be traced back to the
appearance of Gribov copies \cite{Gri78} as merons can be interpreted as
tunneling events between different Gribov vacua \cite{Sobreiro-Sorella}.

However, most of the studies of merons up to now (with the exception of
\cite{prebala}) have been devoted to the $SU(2)$ symmetry group case. In the
present case we will focus on the $SU(N)$-YM theory (for arbitrary values of
$N$) minimally coupled to general relativity (GR). We will be
interested in genuine $SU(N)$ configurations: namely, configurations that are
not trivial embedding of $SU(2)$ into $SU(N)$. This technical detail will be
especially relevant in the analysis of the physical effects of non-embedded
gravitating merons.\footnote{Here it is worth to emphasize that the term
``non-embedded", which will be adopted here, is very common in the literature
on the Skyrme model after the pioneering papers \cite{bala0,Bala1},
where the authors constructed the first numerical examples of genuine $SU(3)$
configurations in the Skyrme model [which are not trivial embeddings of
$SU(2)$ solutions into $SU(3)$].}

The great importance to carefully analyze the coupling of GR with YM theory
arises (at the very least) from two considerations. First of all, there are
situations of high physical interest (such as close to black holes and neutron
stars or in cosmology) in which the coupling of YM theory with GR cannot be
neglected. Moreover, the coupling of topologically non-trivial configurations
in YM theory with GR can be even useful to regularize them. For instance,
merons, which on flat space-times are singular, when coupled to GR can become
regular (see, for instance, \cite{ourmeron1,ourYM1,ourYM2,ourYM3} and references therein).

Many of the results in Einstein-YM are numerical \cite{Bizon,ref1,ref2,ref3,ref4}, and these solutions have been derived
in the case of the $SU(2)$ gauge group. In the Einstein $SU(2)$-YM system
rigorous results are also known \cite{smoller-wasserman} (in-depth analysis of
the $SU(N)$ case can be found in Refs. \cite{volkov,win0,winstanley}).

In the present paper we will construct explicit analytic examples of
non-embedded gravitating merons in the Einstein $SU(N)$-YM theory for arbitrary
values of $N$. However, the main result of the paper is not the construction
of the analytic solutions in itself\footnote{Although in a different form and
with different ansatz, spherical black holes in Einstein $SU(N)$-YM theory
have been already discussed in the literature (see \cite{winstanley} and
references therein).} but rather the non-trivial physical effects which can be
made manifest only with a careful group-theoretical analysis. The solutions
that we will construct below disclose peculiar characteristics of the $SU(N)$
gauge group [which are absent in the $SU(2)$ case] as well as the quite
non-trivial differences between the cases in $D=4$ and $D=5$ dimensions. One
of the interesting features will arise from the analysis of the
\textit{spin-from-isospin\ effect} \cite{JR,tH,goldhaber},
comparing the new configurations with $N>2$ with the usual $N=2$ case.

A similar question about \textquotedblleft genuine $SU(N)$ configurations with
$N\geq3$" in the low energy limit of QCD (which is described by the Skyrme
model \cite{skyrme}) was answered in the seminal works \cite{bala0,Bala1}, and recently in \cite{Sergio,ACLOV,Sergio2}. In
Refs. \cite{bala0,Bala1}, the first \textit{numerical} example
of a \textit{non-embedded} solution representing a \textit{dibaryon} (a bound
state of two baryons) was constructed in the $SU(3)$-Skyrme model [this
numerical construction of non-embedded configurations was extended to the
$SU(N)$-Skyrme model in \cite{BalaSU(N)1}]. Time after, in \cite{ACLOV},
combining the Balachandran ansatz and the generalized hedgehog ansatz with
some known results on the Euler angles for $SU(3)$ \cite{euler1,euler2,euler3}, the first analytical solutions with high
topological charge that describe gravitating dibaryons as well as dibaryons in
flat space-time at finite density were constructed in the Einstein
$SU(3)$-Skyrme model \cite{ACLOV}. These dibaryons are genuine $SU(3)$
features in the sense that they are not trivial embeddings of $SU(2)$ in
$SU(3)$. Finally, very recently, the generalized hedgehog ansatz has been
combined with the Euler parameterization of the $SU(N)$ group describing the
so-called nuclear pasta phases at finite density in the $SU(N)$-Skyrme model
\cite{Sergio,Sergio2}. These solutions are genuine $SU(N)$, due to the
image of $SU(2)$ through the Euler ansatz construction is just a submanifold
but not a subgroup of $SU(N)$, as we will show below. In this sense the map is
not an embedding of $SU(2)$ into $SU(N)$ but just of $S_{3}$ into $SU(N)$
\cite{Dynkin}.

In the present paper, the ansatz proposed in \cite{Sergio} for the
$SU(N)$-Skyrme model will be adapted to the Einstein $SU(N)$-YM case in order
to construct analytical solutions describing non-embedded meronic black holes
(BHs). It is important to highlight that, recently, this ansatz [considering
$\lambda=\lambda(r)$] has allowed the construction of analytical solutions
describing inhomogeneous condensates in the Yang-Mills-Higgs theory in $(2+1)$
dimensions \cite{ourLastFlat} as well as in $(3+1)$ dimensions
\cite{mylastflat}.

The present analysis has three quite non-trivial outcomes. First, one can
tell apart merons BHs from colored BHs using the spin from isospin effect:
while an asymptotically flat meron BH changes the spin of a scalar test field,
a colored black hole does not. This is a very intriguing way to distinguish a
colored BH from a meron BH. Second, using the technology of non-embedded
ansatz in $SU(N)$, \textit{one can generate test fields with arbitrary high
spin}. This is a really powerful result since it allows us to study the dynamics
of higher spin fields without introducing any explicit higher spin field but,
actually, just analyzing the dynamics of a self-interacting scalar field
(charged under the gauge group) living in asymptotically flat $SU(N)$
non-embedded meron BHs (with large enough $N$). It is worthwhile to remind the reader here of the
severe technical problems which are encountered when analyzing the
interactions of higher spin fields related to the Coleman-Mandula theorem and
its generalizations (see \cite{higherspin,higherspin0,higherspin0.1,higherspin0.2,higherspin0.3})
\textquotedblleft preventing" a non-trivial interacting S-matrix in a flat
space for particles with high enough spins.\footnote{We will mention the
relations of the present approach with recent developments in higher spin
field theory in the next sections.}. In this respect, it is worth to note that
there is a class of Higher Spins (HS henceforth) theories in flat space called
Chiral HS which has been constructed in \cite{NEWHS1}. The advantage of Chiral
HS theories is that, at least at one-loop, they avoid the no-go theorems
mentioned above (see \cite{NEWHS2} \cite{NEWHS3}). This approach is based on
\cite{higherspin5.1} \cite{highernew4}.
The present approach provides with a
valid and sound alternative to the analysis of higher spin interactions in
(asymptotically) flat space-times: one can just consider a four-dimensional
renormalizable scalar field theory for a Higgs field (which, consequently, has
quartic vertices) charged under the $SU(N)$ gauge group and living in the
background of a non-embedded $SU(N)$ (gravitating) meron. In the asymptotic
region, due to the presence of the non-embedded meron BH, the scalar field
becomes a higher spin field. Hence, the present construction allows us to study
interacting higher spin fields in asymptotically flat space-times. A further
byproduct of our framework is that the structure of the spin from isospin
effect in $D=4$ is slightly different from the one in $D=5$ dimensions. The
reasons behind this difference will also be discussed.

The paper is organized as follows: in Sec. II we give a brief review of the
Einstein $SU(N)$-YM theory and we present the ansatz that allows us to construct
analytical solutions. In Sec. III we construct BH solutions in $D=4$, and we
study the spin from isospin effect and how higher spin fields can be
generated. In Sec. IV we construct BH solutions in $D=5$ and we compare its
characteristics with those of the $D=4$ case. In Sec. V, using a similar
ansatz, we found an analytic gravitating soliton solution. Sec. VI is
devoted to the conclusions and perspectives.


\section{The Einstein $SU(N)$-Yang-Mills theory}


In this section we make a brief review of the Einstein $SU(N)$-Yang-Mills
theory and also we introduce the general ansatz that allows us to construct
analytical solutions.

\subsection{Field equations}

The action of Einstein $SU(N)$-Yang-Mills theory is given by
\begin{equation}
\label{I}I=\int d^{D} x\sqrt{-g}\biggl(\frac{R-2\Lambda}{\kappa}-\frac
{1}{2e^{2}}\text{Tr}[F_{\mu\nu} F^{\mu\nu}]\biggl) \ ,
\end{equation}
where $R$ is the Ricci scalar, $F_{\mu\nu}=\partial_{\mu}A_{\nu}-\partial
_{\nu}A_{\mu}+i[A_{\mu},A_{\nu}]$ is the field strength of the gauge field
$A_{\mu}$, $\kappa$ is the Newton's coupling constant, $\Lambda$ the
cosmological constant and $e$ is the YM coupling.

Here we use the convention $c=\hbar=1$, Greek indices $\{\mu,\nu,\rho,...\}$
run over the $D$-dimensional space-time with mostly plus signature and Latin
indices $\{a,b,c,...\}$ are reserved for those of the internal space (in the
present paper we will consider the cases $D=4$ and $D=5$).

The YM field equations are
\begin{equation}
\label{EqYM}\nabla_{\nu}F^{\mu\nu}+i[A_{\nu},F^{\mu\nu}] = 0 \ ,
\end{equation}
where $\nabla_{\mu}$ is the Levi-Civita covariant derivative.

The Einstein equations, on the other hand, are given by
\begin{equation}
\label{EqEinstein}R_{\mu\nu}-\frac{1}{2} R g_{\mu\nu} +\Lambda g_{\mu\nu} =
\kappa T_{\mu\nu} \ ,
\end{equation}
with
\begin{equation}
\label{Tmunu}T_{\mu\nu}=\frac{2}{e^{2}}\text{Tr}\biggl(F_{\mu\alpha} F_{\nu
}^{\ \alpha} -\frac{1}{4}g_{\mu\nu} F_{\alpha\beta}F^{\alpha\beta}\biggl) \ ,
\end{equation}
the energy-momentum tensor of the YM field.

\subsection{General ansatz}

We consider a meron-like ansatz for the YM field
\begin{equation}
A_{\mu}\ =\ -i\lambda(x^{\mu})\biggl(U^{-1}\partial_{\mu}U\biggl)\ , \label{A}%
\end{equation}
where $U(x)$ is in a subgroup of $SU(N)$. It is well known that there are many
ways of embedding $SU(2)$ into $SU(N)$. It was Dynkin the first to consider
the classification of such embeddings \cite{Dynkin} (see \cite{bitar} for
details and applications in gauge theory). We choose what is sometimes called
the \textquotedblleft maximal\textquotedblright\ embedding, which is the only
one which gives rise to a irreducible representation of $SU(2)$ of spin
$j=(N-1)/2$ (in agreement with the nomenclature in the Skyrme literature, we
will call these configurations ``non-embedded"). We may parameterize it in
terms of the generalized Euler angles as follows
\begin{equation}
U\ =\ e^{-iF_{1}(x^{\mu}) T_{3}}e^{-iF_{2}(x^{\mu})T_{2}}e^{-iF_{3}(x^{\mu
})T_{3}}\ , \label{U}%
\end{equation}
where the matrices $T_{a}$ are explicitly given by
\begin{align}
T_{1}=  &  \frac{1}{2}\sum_{j=2}^{N}\sqrt{(j-1)(N-j+1)}(E_{j-1,j}%
+E_{j,j-1})\ ,\label{eq:Ts}\\
T_{2}=  &  \frac{i}{2}\sum_{j=2}^{N}\sqrt{(j-1)(N-j+1)}(E_{j-1,j}%
-E_{j,j-1})\ ,\\
T_{3}=  &  -\sum_{j=1}^{N}\biggl(\frac{N+1}{2}-j\biggl)E_{j,j}\ , \label{T3}%
\end{align}
with
\begin{equation}
(E_{i,j})_{mn}=\delta_{im}\delta_{jn}\ . \label{eqts2}%
\end{equation}
They are chosen so that the following relations are satisfied:
\begin{equation}
\lbrack T_{a},T_{b}]=i\epsilon_{abc}T_{c}\ ,\qquad\text{Tr}(T_{a}T_{b}%
)=\frac{N(N^{2}-1)}{12}\delta_{ab}\ . \label{traces}%
\end{equation}
It is worthwhile to emphasize that the above generators are an irreducible
representation of $SU(2)$, which is not true for all embbedings \cite{euler1,euler2,euler3}. In the case of $SU(3)$, for instance, one may
take one half of the first three Gell-Mann matrices as generators of $SU(2)$,
which form a spin $1/2$ representation of $SU(2)$. However, it is not
irreducible, because its three $3\times3$ matrices have zeros everywhere
except for their $2\times2$ first blocks, where the spin matrices are
embedded. The above $T_{a}$ matrices, on the contrary, form the spin-$j$
irreducible representation of $SU(2)$, with $j=(N-1)/2$. This may be seen
directly from the diagonal element \eqref{T3}, or by noting that
\begin{align}
\label{Tcua}\left(  \overrightarrow{T}\right)  ^{2}  &  =\sum_{a=1}^{3}%
T_{a}T_{a}=\sigma\left(  N\right)  \mathbf{1\ ,}\\
\sigma\left(  N\right)   &  =\frac{(N^{2}-1)}{4}=j(j+1)\ . \label{sigma}%
\end{align}
Picking the irreducible representation of $SU(2)$ for all values of $N$
implies that for every $N$ we are using a representation with different spin.
This means that $\left(  \overrightarrow{T}\right)  ^{2}$\ (which will play an
important role to define the ``square of the total angular momentum operator")
depends on $N$. One can see that $\sigma(N)$ grows with $N^{2}$ so that, for
the irreducible embedding ansatz presented here, the total angular momentum
will also grow with $N$ (as it will be discussed in the next sections).

\subsection{A short review on merons}

Classic results on gravitating merons and their physical applications in the
case of Einstein-YM theory with the $SU(2)$ gauge group are in
\cite{teitelboim,witten,old1,old2,old3,old4,old5,old6,old7,old8,old9}.
\footnote{It is interesting to note that in \cite{teitelboim} the authors
constructed the first example of a $SU(2)$ meron black hole. However, the
concept of meron was invented after such black hole was constructed. That is
why the authors of \cite{teitelboim} do not mention the connection with
merons.}

A meron can always be brought in the following form:
\begin{align}
A_{\mu} = -i\lambda\left(  U^{-1}\partial_{\mu}U \right)  \ , \quad\lambda
\neq0,1 \ , \label{meronsansatz1}%
\end{align}
which is proportional to a pure gauge term without being, of course, a pure
gauge configuration. Therefore the existence of merons is an
\textit{intrinsically non-Abelian feature}. The first example on flat
space-time was constructed by de Alfaro, Fubini and Furlan in Ref.
\cite{merons1}, and it has $\lambda=1/2$. Although, in principle, $\lambda$
could take any value different from zero and one, here we will show that even
in the case of the $SU(N)$ gravitating meron $\lambda=1/2$ is indeed a special value.

The field strength $F_{\mu\nu}$ of the meron in Eq. (\ref{meronsansatz1}) is
proportional to the commutator,%
\begin{equation}
F_{\mu\nu}=-i\lambda\left(  \lambda-1\right)  \left[  U^{-1}\partial_{\mu
}U,U^{-1}\partial_{\nu}U\right]  \,. \label{curvamerons}%
\end{equation}
Recently\footnote{Using a strategy developed originally to analyze the Skyrme
model (see \cite{Sergio2,56,58b,56a,56b,56c,56d,56e,gaugsk,LastUS1}).} in
\cite{ourmeron1,ourYM1,ourYM2,ourYM3}, it has been
possible to analyze explicitly the physical effects generated by $SU(2)$ meron
BHs. In particular, it has been shown that the asymptotically flat case is a
very interesting arena to implement the usual spin from isospin effect without
worrying about the singularities associated to the meron (which are hidden
behind the BH horizon). In the present paper, we will ask the following questions:

\begin{enumerate}
\item Is the Einstein $SU(N)$-YM case physically different from the already
known $SU(2)$ case?

\item Are there genuine $SU(N)$ configurations which are absent in the $SU(M)$
case with $M<N$?

\item Which are the physical effects associated to these genuine $SU(N)$ configurations?
\end{enumerate}

The above interesting questions can be answered in a very elegant way
combining the group theoretical tools developed in Refs. \cite{euler1,euler2,euler3}, both with the idea of non-embedded ansatz
developed in \cite{bala0,Bala1}, as well as with the recent results in
\cite{Sergio,ACLOV}.


\section{Black holes in $D=4$}


In this section we construct meron BHs in the Einstein $SU(N)$-YM theory in
$D=4$.

\subsection{Analytic meron black hole solutions}

We impose spherical symmetry considering the metric
\begin{equation}
ds^{2}=-f(r)dt^{2}+\frac{1}{f(r)}dr^{2}+r^{2}\left(  d\theta^{2}+\sin
^{2}{\theta}d\phi^{2}\right)  \ . \label{metric4}%
\end{equation}
The meron in Eq. \eqref{A} that satisfies identically the complete set of YM
equations in Eq. \eqref{EqYM} is given by
\begin{equation}
\label{Fs}F_{1}(x^{\mu})=-\phi\ ,\quad F_{2}(x^{\mu})=2\theta\ ,\quad
F_{3}(x^{\mu})=\phi\ ,
\end{equation}
together with the particular value of $\lambda$ mentioned above,
\begin{equation}
\label{1/2}\lambda\ =\ \frac{1}{2}\ .
\end{equation}
From the Einstein equations in Eq. \eqref{EqEinstein}, we obtain for the
metric function $f(r)$ the following expression
\begin{align}
f(r)\ =\  &  1-\frac{2m}{r}-\frac{\Lambda}{3}r^{2}+\frac{8\lambda^{2}%
(\lambda-1)^{2}\kappa}{e^{2}r^{2}} \frac{(N-1)N(N+1)}{6}\label{f(r)}\\
=\  &  1-\frac{2m}{r}-\frac{\Lambda}{3}r^{2}+\frac{\kappa}{2e^{2}r^{2}%
}T_{\text{N}}\ ,\nonumber
\end{align}
with $T_{\text{N}}=\frac{(N-1)N(N+1)}{6}$ as the Tetrahedral numbers for
$N=2,3,..$.

It turns out that the meron in this case is just the Wu-Yang monopole, whose
singularity is dressed under the BH horizon. In fact,
\begin{equation}
A^{i}=-\frac{1}{r^{2}}\epsilon^{ija}x_{j}T_{a}\ , \label{WY}%
\end{equation}
where, $(x_{1},x_{2},x_{3})=r(\sin\theta\cos\phi,\sin\theta\sin\phi,\cos
\theta)$. The above solution has exactly the same form as the one in Minkowski
space-time, but be aware that the $x^{i}$ are only asymptotically the
Cartesian coordinates of flat space. It is a straightforward computation to
check that twice the Wu-Yang monopole field in Eq. \eqref{WY} gives vanishing
field strenght, that is, as pure gauge as expected for a meron with
$\lambda=1/2$. Now, if one performs a gauge transformation using a group
element of the form \eqref{U}, with
\begin{equation}
F_{1}(x^{\mu})=-\phi\ ,\quad F_{2}(x^{\mu})=-\theta\ ,\quad F_{3}(x^{\mu
})=\phi\ ,
\end{equation}
then the YM potential transforms to the \textquotedblleft Dirac gauge"
\begin{equation}
A=(1-\cos\theta)d\phi T_{3}\ .
\end{equation}
This potential has a Dirac string singularity at $\theta=0$, and the field
strength is given by
\[
F_{\mu\nu}=f_{\mu\nu}T_{3}\ ,
\]
where $f_{\mu\nu}$ is the field of the Dirac monopole, with $f_{\theta\phi
}=\sin\theta$, the only non-vanishing component. The field is effectively
Abelian, and its contribution to the action in Eq. \eqref{I} is
\[
\frac{1}{2e^{2}}\int d^{4}x\sqrt{-g}f_{\mu\nu}f^{\mu\nu}\text{Tr}[T_{3}%
^{2}]=\frac{N(N^{2}-1)}{24e^{2}}\int d^{4}x\sqrt{-g}f_{\mu\nu}f^{\mu\nu}\ ,
\]
where we have used Eq. \eqref{traces}. This means that, in fact, the effective
coupling constant $Q$ is given by
\begin{equation}
Q^{2}=\frac{12e^{2}}{N(N^{2}-1)}\ . \label{Q2}%
\end{equation}
The resulting metric is precisely the Reissner-Nordstr\"{o}m metric in
(anti-)de Sitter space-time with unitary magnetic charge,
\begin{equation}
g=\frac{1}{Q}=\frac{N(N^{2}-1)}{12e^{2}}. \label{g2}%
\end{equation}
Note that the monopole in the Dirac gauge is not of the form
\eqref{meronsansatz1}. Its double is not pure gauge. Actually, it may be
multiplied by any constant to get a monopole solution with any magnetic
charge. However, if the magnetic charge is not unitary, then we will not be
able to perform a gauge transformation that takes it to the Wu-Yang form, that
is, it will not be a meron anymore. Indeed, the gauge transformation from the
meronic configuration to the Abelian Dirac monopole is singular at the origin
(see the discussion on pages 13 and 14 of \cite{SuperLastM}). Since two gauge
potentials are gauge equivalent if and only if there is a \textit{proper gauge
transformation} (namely, a \textit{smooth gauge transformation which is also
well behaved at infinity}\footnote{Well behaved at infinity means that the
group-valued element $U$ which generates such gauge transformation must
approach the center of the gauge group at spatial infinity: see the discussion
in \cite{ilderton}. Note that the group element of the form \eqref{U} [with
$F_{1}(x^{\mu})=-\phi$, $F_{2}(x^{\mu})=-\theta$, $F_{3}(x^{\mu})=\phi$]\ not
only is singular at the origin but also does not approach the center of
$SU(2)$ (which is $\pm\mathbf{1}_{2\times2}$) at spatial infinity.}) from one
configuration to the other, one can conclude that the present meronic
configuration and the Dirac monopole are not gauge equivalent. Note also that
if one would not define gauge equivalence using proper gauge transformations
one would arrive at absurd conclusions such as that the (anti-)de Sitter
space-time in $(2+1)$ dimensions is the same as the Bañados-Teitelboim-Zanelli black hole (as these
two configurations are connected by an improper gauge transformation).

Even though the above solution is well known, there is an interesting feature
arising from the dependence of the effective charge $g$ with $N$ as seen in
Eq. \eqref{g2}. If the cosmological constant $\Lambda$ is positive, then for
a horizon to exist the magnetic (or electric) charge must satisfy
$g^{2}<(4\Lambda)^{-1}$. Therefore, these merons cease to exist for big enough
$N$. There are also bound for the mass. If the cosmological constant vanishes,
for instance, then for a horizon to dress the singularity the mass must be
such that $M^{2}>g^{2}$. Therefore, as $N$ grows, the mass of the merons are
forced to grow as well.

Obviously, spherically symmetric BHs in the Einstein $SU(N)$-YM theory have
been already discussed in depth in the literature (see, for instance,
\cite{volkov,win0,winstanley,su(n)BH1,su(n)BH2,su(n)BH3} and references therein). In fact, the idea of
the present construction (using an explicit ``non-embedded" ansatz for the
meronic field) is that it discloses in a very neat way the fact that the spin
from isospin effect depends actually on ``the $N$" of the gauge group $SU(N)$,
so that the interactions of test scalar fields [charged under $SU(N)$] with
the gravitating merons discussed here can generate fields of arbitrary high
spin (if $N$ is large enough). This fact has not been noticed before (to the
best of our knowledge) and is a novel outcome of our technique.

\subsection{About colored black holes}

It is well known that the Einstein-YM theory admits spherically symmetric BHs
solutions with a non-Abelian hair (see \cite{colored1,colored2,colored3} and references therein) in which the non-Abelian electric and
magnetic fields decay too fast to give rise to charges. Despite their
instability \cite{colored4,colored5}, the very important role of such
non-Abelian hairy BHs (especially in the application of holography) cannot be
underestimated \cite{colored5.1,colored5.2}. Here we want just to
emphasize that these BHs can be written very easily using the present
approach. We will consider the following metric
\begin{equation}
ds^{2}=-f(r)dt^{2}+\frac{1}{h(r)}dr^{2}+r^{2}d\theta^{2}+r^{2}\sin^{2}{\theta
}d\phi^{2}\ ,
\end{equation}
together with a radial profile for the YM field, namely
\[
A_{\mu}\ =\ -i\lambda(r)(U^{-1}\partial_{\mu}U)\ ,
\]
and
\begin{equation}
F_{1}(x^{\mu})=-\phi\ ,\quad F_{2}(x^{\mu})=2\theta\ ,\quad F_{3}(x^{\mu
})=\phi\ .
\end{equation}
Of course, $\lambda=1/2$, would give the meron BH, while hairy colored BHs
must be found numerically. The YM equations are reduced to the following
equation for the profile
\begin{equation}
\lambda^{\prime\prime}+\frac{(fh)^{\prime}}{2fh}\lambda^{\prime}%
-\frac{2\lambda(\lambda-1)(2\lambda-1)}{r^{2}h}\ =\ 0\ .
\end{equation}
On the other hand, the components of the energy-momentum tensor are
\begin{align*}
T_{tt}\ =\  &  4T_{\text{N}}\times\frac{f}{e^{2}r^{4}}(2\lambda^{2}%
-4\lambda^{3}+2\lambda^{4}+r^{2}h\lambda^{\prime2})\ ,\\
T_{rr}\ =\  &  4T_{\text{N}}\times-\frac{1}{e^{2}hr^{4}}(2\lambda^{2}%
-4\lambda^{3}+2\lambda^{4}-r^{2}h\lambda^{\prime2})\ ,\\
T_{\theta\theta}\ =\  &  4T_{\text{N}}\times\frac{2}{e^{2}r^{2}}%
(\lambda-1)^{2}\lambda^{2}\ ,\\
T_{\phi\phi}\ =\  &  \sin^{2}\theta T_{\theta\theta}\ ,
\end{align*}
while the components of the Einstein tensor (with cosmological constant) are
given by
\begin{align*}
G_{tt}+\Lambda g_{tt}=  &  \frac{f}{r^{2}}(1-h-rh^{\prime})-\Lambda f\ ,\\
G_{rr}+\Lambda g_{rr}=  &  \frac{1}{r^{2}fh}(fh-f+rhf^{\prime})+\Lambda
\frac{1}{h}\ ,\\
G_{\theta\theta}+\Lambda g_{\theta\theta}=  &  \frac{r}{4f^{2}}%
\biggl(f[rf'h'+2h(f'+rf'')]+2f^2h'-rhf'^2  \biggl)    +\Lambda r^{2}\ ,\\
G_{\phi\phi}+\Lambda g_{\phi\phi}=  &  \sin^{2}\theta(G_{\theta\theta}+\Lambda
g_{\theta\theta}) \ .
\end{align*}
This equations system (where $N$ only enters as an overall factor in the
energy-momentum tensor) has been already analyzed, so that the known numerical
solutions of the references mentioned above can be adapted to the present case.

Here we only want to mention that the key difference between meron BHs and
colored BHs appears in the Klein-Gordon equation%
\begin{equation}
\label{Klein}(\Box-m^{2}) \mathbf{\Phi}\ =\ 0\ , \qquad\Box=D_{\mu}D^{\mu}\ ,
\end{equation}
for a scalar field $\mathbf{\Phi}$ charged under the gauge group. In the
asymptotically flat case, the terms that should give rise to the spin from
isospin effect [which are $g^{\mu\nu}A_{\mu}A_{\nu}\mathbf{\Phi}$\ and
$g^{\mu\nu}\left(  A_{\mu}\right)  \nabla_{\nu}\mathbf{\Phi}$] decay faster
than in the case of the meron BH, so that, in the asymptotic region of the
colored BHs, such terms are unable to form the contribution ``$\left(
\overrightarrow{J}\right)  ^{2}/r^{2}$" (which will be discussed in the next
section) needed to transform Bosons into Fermions (and vice versa).

\subsection{Gravitational spin from isospin effect in $SU(N)$}

In general, the presence of a background field breaks the natural symmetries
of a theory. For instance, the $SU(N)$ Klein-Gordon or Dirac equations, in
which the Yang-Mills field is explicitly given, will break rotational
invariance (unless the given field is spherically symmetric). However, there
are situations in which the field is indeed symmetric, but the corresponding
gauge potential, which appears in the equations, is not. In that case, the
orbital angular momentum $\overrightarrow{l}$ will not be a symmetry
generator. However, it is possible to compensate the lack of invariance of the
potential under spatial rotations with an appropriate gauge rotation. For
example, the potential in Eq. \eqref{WY} is not invariant under rotations.
However, if one performs the same $SU(2)$ gauge rotation to both space-time
indices and internal indices, then the symmetry is recovered. The operator
that generates such a transformation is
\begin{equation}
\label{J}\overrightarrow{J}=\overrightarrow{l}+\overrightarrow{T} \ ,
\end{equation}
where the vector $\overrightarrow{T}$ is formed by the generators of the
non-embedded subgroup of $SU(N)$ defined in Eqs. (\ref{eq:Ts})--(\ref{eqts2}),
while $\overrightarrow{l}$ is the usual orbital angular momentum operator.
Hence, $\overrightarrow{J}$ should be considered as the total angular
momentum of the system.

It is precisely this \textit{spherical symmetric up to an internal\ rotation}
which gives rise to the \textit{Jackiw-Rebbi-Hasenfratz-'t Hooft} mechanism,
or \textquotedblleft spin form isospin\textquotedblright\ effect \cite{JR,tH}, according to which the excitations of a Bosonic field charged under
$SU(2)$ around a background gauge field with the above characteristics behave
as Fermions.\footnote{An effect which is very similar to the
\textit{Jackiw-Rebbi-Hasenfratz-'t Hooft }mechanism occurrs for Skyrmions
\cite{skyrme} (for a detailed review, see \cite{manton}). Indeed, the
excitations around the Skyrme soliton with winding number equal to one can
behave as Fermions.} We are interested here in the case of $SU(N)$, in which
the meron solution discussed in the previous section will do the same trick. A
quick way to derive the spin from isospin phenomena is to analyze the
Klein-Gordon equation in Eq. \eqref{Klein} for a scalar field $\mathbf{\Phi}$
(which will be assumed to belong to the fundamental representation) charged
under $SU(N)$, being in this case $\nabla_{\mu}$ the Levi-Civita covariant
derivative corresponding to the metric in Eqs. \eqref{metric4} and
\eqref{f(r)}, and $A_{\mu}$ is the $SU(N)$ meron gauge potential in Eqs.
\eqref{A}, \eqref{U} and \eqref{Fs}. For the present purpose, it is enough to
restrict us to the static case, set $\Lambda=0$ and to explore the asymptotic
region, where the metric is Minkowski. We also set $m=0$, so that Eq.
\eqref{Klein} becomes
\begin{equation}
(\nabla_{i}+iA_{i})(\nabla^{i}+iA^{i})\mathbf{\Phi}=\Big(\nabla^{2}%
+2iA_{i}\nabla^{i}+i(\nabla^{i}A_{i})-A_{i}A^{i}\Big)\mathbf{\Phi}\ .
\label{kg}%
\end{equation}
The first term in Eq. \eqref{kg} is the Laplacian,
\[
\nabla^{2}\mathbf{\Phi}=\frac{1}{r^{2}}\Big[\partial_{r}(r^{2}\partial
_{r}\mathbf{\Phi})-\vec{L}^{2}\mathbf{\Phi}\Big]\ ,
\]
where
\[
\vec{L}=-i\vec{r}\times\vec{\nabla}\ ,
\]
is the orbital angular momentum operator. Using Eq. \eqref{WY}, the second
term in Eq. \eqref{kg} is
\[
2iA_{i}\nabla^{i}=\frac{2i}{r^{2}}T_{a}\epsilon^{aji}x_{j}\nabla_{i}%
=-2T_{a}L_{a}\ .
\]
The third term vanishes because $\nabla^{i}A_{i}=0$, as one may verify
directly. Finally, for the last term,
\[
-A_{i}A^{i}=-\frac{1}{r^{4}}(r^{2}\delta^{ab}-x^{a}x^{b})T_{a}T_{b}=-\frac
{1}{r^{2}}[\vec{T}^{2}-(\hat{r}\cdot\vec{T})^{2}]\ ,
\]
where $\hat{r}\cdot\vec{T}=x_{a}T_{a}/r$ is the projection of $\overrightarrow
{T}$ along the direction of $\vec{r}$. Putting all together, Eq. \eqref{kg}
turns out to be
\begin{align}
0  &  =\frac{1}{r^{2}}\partial_{r}(r^{2}\partial_{r}\mathbf{\Phi})+\frac
{1}{r^{2}}\Big(-\vec{L}^{2}-2T_{a}L_{a}-\vec{T}^{2}+(\hat{r}\cdot\vec{T}%
)^{2}\Big)\mathbf{\Phi}\label{KGd4}\\
&  =\frac{1}{r^{2}}\partial_{r}(r^{2}\partial_{r}\mathbf{\Phi})-\frac{1}%
{r^{2}}\Big(\vec{J}^{2}-(\hat{r}\cdot\vec{T})^{2}\Big)\mathbf{\Phi
}\ .\nonumber
\end{align}
Here $\overrightarrow{J}$ is the total angular momentum in Eq. \eqref{J}. We
see that it forms in the Klein-Gordon equation, supplementing the orbital part
as it should. Therefore, one can generate higher spin fields in asymptotically
flat space-times using test scalar fields (charged under the gauge group)
living in the $SU(N)$ meron BHs constructed in the previous subsections.

\subsection{Higher spin fields from non-embedded ansatz in $D=4$}

The classic results in \cite{higherspin,higherspin0,higherspin0.1,higherspin0.2,higherspin0.3}, showed that,
under ``normal" circumstances, in flat space-times one cannot formulate a
consistent quantum field theory with massless particles with spins greater than two. The same
approach also suggests similar negative results in asymptotically flat
space-times. Soon after these original references, some positive partial
results on how to define consistent (cubic) interactions between higher spins
fields were obtained in \cite{higherspin5.1,higherspin5,higherspin5.2}. However, the problem to define consistent renormalizable
interactions between higher spin fields on (asymptotically) flat space-times
remained. A situation with negative cosmological constants (due to its role
as an effective infrared cutoff) was disclosed in \cite{higherspin3,higherspin3.1} (an in-depth analysis of the current situation can be
found in \cite{higherspin2,higherspin4,higherspin1,highernew,highernew1,highernew2,highernew3} and
references therein). It is worth
mentioning that there are also no-go theorems also in AdS (see for instance
\cite{highernew3.1} \cite{highernew3.2} \cite{highernew3.3}): the present
proposal, to be described here below, can be very useful also in order to
avoid these AdS no-go theorems.

To the best of authors' knowledge, the only
well-established case (so far) in which it is possible to define a consistent
interaction in four-dimensional (asymptotically) flat space-times is the cubic
vertex (see, for a modern perspective, \cite{highernew4,highernew5,highernew6} and references therein). In particular, in those references,
a complete classification of the possible cubic vertices has been performed.
It is worth to emphasize that, within their approach, the spectrum is
reducible and consist of propagating massless particles with spin $s$, $s-2$,
$s-4$, ... and so on. Consequently, this modern formulation is different from
\cite{higherspin5}, in which case the field equations describe a single
massless degree of freedom of a particile with spin $s$.

In this sense, the spin from isospin effect corresponding to the non-embedded
gravitating merons constructed in the previous sections is more similar to
\cite{higherspin5} rather than to the modern references mentioned above. The
reason is that with the choice of the generators in Eqs. (\ref{eq:Ts}),
(\ref{T3}) and (\ref{eqts2}) one gets an irreducible representation of $SO(3)$
of spin $j=(N-1)/2$. Hence, due to the \textit{conversion of isospin into
spin} (see \cite{monopolescattering1,monopolescattering2}) a scalar
field charged under the gauge group $SU(N)$ becomes a field of spin
$j=(N-1)/2$. One way to see this (which has been already discussed in the
previous sections) is that the ansatz for the gauge field in Eqs. (\ref{A}),
(\ref{U}), (\ref{eq:Ts}), (\ref{T3}) and (\ref{eqts2}) is not spherically
symmetric, but the lack of spherical symmetry can be compensated by an
internal rotation of spin $j=(N-1)/2$ so that the ``true" angular momentum
operator acting on such a scalar field corresponds to a spin-$j$ field.

Now, if one wants to consider interactions one can analyze the well-known
(renormalizable in $D=4$) scalar field Lagrangian for the Higgs field charge
under the $SU(N)$ gauge group with a quartic Higgs potential whose field
equations and Lagrangian read, respectively,
\begin{equation}
g^{\mu\nu}\left(  \nabla_{\mu}+iA_{\mu}\right)  \left(  \nabla_{\nu}+iA_{\nu
}\right)  \mathbf{\Phi}=-\gamma\left(  v^{2}-|\mathbf{\Phi}|^{2}\right)
\mathbf{\Phi}\ , \label{equint}%
\end{equation}%
\begin{equation}
I\left[  \mathbf{\Phi}\right]  =\frac{1}{4}\int d^{4}x\sqrt{-g}%
\biggl(\text{Tr}[D_{\mu}\mathbf{\Phi}D^{\mu}\mathbf{\Phi}]-\gamma\left(
v^{2}-|\mathbf{\Phi}|^{2}\right)  ^{2}\biggl)\ , \label{equint2}%
\end{equation}%
\[
\mathbf{\Phi}^{2}=-\frac{1}{2}\text{Tr}[\mathbf{\Phi\Phi}]\ .
\]
The above theory is renormalizable in $D=4$ and the corresponding Feynman
rules in coordinates space can be defined in the usual way (taking care of the
non-trivial background). In order to display the interplay between the
vertices and the spin of $\mathbf{\Phi}$, one can expand explicitly in terms
of eigenfunctions $\mathbf{\Phi}$ of $\overrightarrow{J}^{2}$ and $\hat
{r}\cdot\vec{T}$. Clearly, being the original theory $I\left[  \mathbf{\Phi
}\right]  $ well defined in $D=4$, the interaction vertices will be well
defined as well, and, since the field $\mathbf{\Phi}$ acquires a spin
$j=(N-1)/2$ due to the background, one can interpret the usual Feynman rules
as Feynman rules for spin $(N-1)/2$ fields. The original no-go theorems
\cite{higherspin,higherspin0,higherspin0.1,higherspin0.2,higherspin0.3}, are avoided since the presence of
the gravitating meron breaks the symmetry of the vacuum and changes the
topology of space-time. Thus, as long as the backreaction of $\mathbf{\Phi}$
on the background can be neglected, in principle this construction works. Of
course, there are severe technical complications to implement this program in
practice due to the fact that the non-trivial background prevents one from
finding easily the propagators in Fourier space. We hope to return to this
interesting issue in a future work.


\section{Black holes in $D=5$}


In this section we construct meron BHs in the Einstein $SU(N)$-YM theory in
$D=5$.

\subsection{Analytic meronic black hole solutions}

We consider a five-dimensional, spherically symmetric, space-time ansatz:
\begin{equation}
\label{metric5}ds^{2}=-f(r)^{2} dt^{2} + \frac{1}{f(r)^{2}}dr^{2}+\frac{r^{2}%
}{4}\biggl( d\gamma^{2} +d\theta^{2} +d\phi^{2} +2\cos{\theta} d\gamma
d\phi\biggl) \ ,
\end{equation}
together with the YM field given by Eqs. \eqref{A} and \eqref{U}, with
\begin{equation}
\label{Fs2}F_{1}(x^{\mu}) = -\phi\ , \quad F_{2}(x^{\mu}) = -\theta\ , \quad
F_{3}(x^{\mu}) = -\gamma\ ,
\end{equation}
\begin{equation}
\lambda= \frac{1}{2} \ .
\end{equation}
These fields satisfy the YM equations. They correspond to a $D=5$ meron, an
analog of the $D=4$ case described in the previous section. The Einstein
equations may be explicitly solved:
\begin{align}
f(r)^{2} \ = \  &  1 - \frac{2m}{r^{2}} - \frac{\Lambda}{6}r^{2} - \frac{2}{3}
\times24 (\lambda-1)^{2}\lambda^{2} \frac{\kappa\log(r)}{e^{2} r^{2}}
\frac{(N-1)N(N+1)}{6}\nonumber\\
\ = \  &  1 - \frac{2m}{r^{2}} - \frac{\Lambda}{6}r^{2} - \frac{\kappa\log
(r)}{e^{2} r^{2}} T_{\text{N}} \ . \label{BH5}%
\end{align}
Here $T_{\text{N}} = \frac{(N-1)N (N +1)}{6}$ are the tetrahedral numbers. The
constant $\lambda$ has been left arbitrary so one can see that when the YM
field is pure gauge, $\lambda=1$, the metric reduces to Schwarzschild-(anti-)
de Sitter in $D=5$.

\subsection{Gravitational spin from isospin effect in $SU(N)$transo}

As in the previous section, in order to study the spin from isospin effect, we
will analyze the Klein-Gordon equation in Eq. \eqref{Klein} in $D=5$ for a
scalar field $\mathbf{\Phi}$ charged under $SU(N)$, with $\nabla_{\mu}$ the
Levi-Civita covariant derivative corresponding this time to the metric in Eqs.
\eqref{metric5} and \eqref{BH5}. Here $A_{\mu}$ is the $SU(N)$ meron gauge
potential in Eqs. \eqref{A}, \eqref{U} and \eqref{Fs2}.

In $D=5$ the orbital angular momentum is given, in Cartesian coordinates, by
\[
\mathcal{L}_{AB}=-i(x_{A}\partial_{B}-x_{B}\partial_{A})\ ,
\]
where $x^{A}$, $A=1,\ldots,4$ are the spatial indices. They satisfy the
$SO(4)$ algebra. Because $SO(4)=SO(3)\times SO(3)$, the above generators may
be divided into two sets, each satisfying the $SO(3)$ algebra. Explicitly,
\begin{equation}
L_{a}^{\pm}=\epsilon_{a}{}^{bc}\mathcal{L}^{bc}\pm\mathcal{L}_{4a}%
\ ,\qquad\lbrack L_{a}^{\pm},L_{b}^{\pm}]=i\epsilon^{c}{}_{ab}L_{c}^{\pm
}\ ,\qquad\lbrack L_{a}^{+},L_{b}^{-}]=0\ , \label{LL}%
\end{equation}
where $a,b=1,2,3$. We call $L_{a}^{+}$, $L_{b}^{-}$ the right and left angular
momentum, respectively. It is useful to write these generators in the spherical
coordinates of the $3$-sphere defined in the metric \eqref{metric5}. For
example, the right angular momentum is given by
\begin{align}
L_{1}^{+}=  &  i\biggl(\cos\gamma\cot\theta\partial_{\gamma}+\sin
\gamma\partial_{\theta}-\frac{\cos\gamma}{\sin\theta}\partial_{\phi
}\biggl)\ ,\\
L_{2}^{+}=  &  i\biggl(\sin\gamma\cot\theta\partial_{\gamma}-\cos
\gamma\partial_{\theta}-\frac{\sin\gamma}{\sin\theta}\partial_{\phi
}\biggl)\ ,\\
L_{3}^{+}=  &  -i\partial_{\gamma}\ .
\end{align}
In this form, the generators are well defined not only in Minkowski space but
also in the BH geometry in Eq. \eqref{metric5}. In terms of these, the
D'Alambert operator is
\begin{align*}
\Box &  =-\frac{1}{f^{2}}\partial_{t}^{2}+\frac{1}{r^{3}}\partial_{r}%
(r^{3}f^{2}\partial_{r})-\frac{1}{r^{2}}\frac{1}{2}\mathcal{L}_{AB}%
\mathcal{L}^{AB}\\
&  =-\frac{1}{f^{2}}\partial_{t}^{2}+\frac{1}{r^{3}}\partial_{r}(r^{3}%
f^{2}\partial_{r})-\frac{2}{r^{2}}\big[(\vec{L}^{+})^{2}+(\vec{L}^{-}%
)^{2}\big]\ .
\end{align*}
As in the $D=4$ case, we now consider the Klein-Gordon equation for a scalar
field $\mathbf{\Phi}$ in the background of the right-handed meron,
\begin{equation}
(\Box+i\nabla_{\mu}A^{\mu}+2iA^{\mu}\nabla_{\mu}-A^{\mu}A_{\mu}-m^{2}%
)\mathbf{\Phi}=0\ . \label{kg2}%
\end{equation}
Substituting the explicit expressions for $A_{\mu}$ and $g_{\mu\nu}$ given by
Eqs. \eqref{A}, \eqref{U}, \eqref{metric5}--\eqref{BH5}, Eq. \eqref{kg2} takes
the form
\begin{equation}
\biggl(-\frac{1}{f^{2}}\partial_{t}^{2}+\frac{1}{r^{3}}\partial_{r}(r^{3}%
f^{2}\partial_{r})-\frac{1}{r^{2}}\Big(2(\vec{J}^{+})^{2}+2(\vec{J}^{-}%
)^{2}-\sigma(N)\mathbf{1}\Big)-m^{2}\biggl)\mathbf{\Phi}=0\ , \label{kg22}%
\end{equation}
where $\mathbf{1}$ is the $N\times N$ identity matrix, $\sigma(N)$ is given in
Eq. \eqref{sigma} and
\[
J_{a}^{+}=L_{a}^{+}+T_{a}\ ,\qquad J_{a}^{-}=L_{a}^{-}\ .
\]
From this equation we see that the angular momentum is given by the pair
$J_{a}^{+}$, $J_{a}^{-}$ which, besides the orbital part $L_{a}^{+}$,
$L_{a}^{-}$, has a contribution from the generators of $SU(N)$. In this case,
only the right angular momentum $J_{a}^{+}$ gets shifted. Of course, there is
nothing special about the right angular momentum. A second solution of the
Yang-Mills-Einstein system exists which shifts the left angular momentum
$J^{-}_{a}$ instead. It is obtained by replacing the group element $U$ of the
above solution by by $U^{-1}$. The metric \eqref{f(r)} and the meron form
\eqref{A} of the gauge field are the same.

Note also that, in addition to the angular momentum, the expression
multiplying $r^{-2}$ in Eq. \eqref{kg22} contains a term proportional to the
identity. This is much simpler than the $D=4$ case, where the extra term is
$(\hat{r}\cdot\vec{T})^{2}$, as seen in Eq. \eqref{KGd4}. The reason behind
this reduction (similar to what happens for the BH in \cite{ourmeron1}), lies
in the term $A_{\mu}A^{\mu}$ in Eq. \eqref{kg2}, which in the $D=5$ case,
turns out to be proportional to $(\overrightarrow{T})^{2}$. Then, the spin of
the particles becomes exactly $\sigma_{N}$ according to Eq. \eqref{Tcua}.

There is another important difference between the black hole solutions in
$D=4$ and $D=5$ presented above, namely, the first has vanishing topological
charge while the latter has a finite one. In fact, consider the following
standard definition of the topological charge,
\[
B=\frac{1}{24\pi^{2}}\int_{\Sigma}\rho_{\text{B}}\ ,\qquad\rho_{\text{B}%
}=\epsilon^{ijk}\text{Tr}[\mathcal{L}_{i}\mathcal{L}_{j}\mathcal{L}_{k}]\ ,
\]
where $\Sigma$ is any three-dimensional spatial surface defined by $t=const$
and $r=const$\ while
\begin{equation}
\mathcal{L}_{\mu}=U^{-1}\partial_{\mu}U=\Omega_{\mu}^{a}T_{a}\ , \label{left}%
\end{equation}
are the Maurer-Cartan form components, $\Omega_{\mu}^{a}$ are the
left-invariant 1-forms components of an element $U(x)\in SU(N)$ parameterized
as in Eq. \eqref{U}. Note that a necessary (but not sufficient) condition for
having a non-zero topological charge is that the functions $F_{i}$ in Eq.
\eqref{U} must be independent. This effectively occurs in the case of the BH
in $D=5$ considered above, where each function depends linearly on a different
coordinate of the $3$-sphere [see Eq. \eqref{Fs2}], and it is possible to
verify that $B\neq0$ on $\Sigma$ by integrating into the ranges of the
coordinates in Eq. \eqref{metric5}. On the other hand, in the case of the
meron BH in $D=4$, the functions $F_{i}$ are not independent [see Eq.
\eqref{Fs}], and it is direct to check that $\rho_{B}=0$ identically.


\section{Gravitating soliton}


In this section we present an analytic self-gravitating soliton solution in
$D=4$. Although this configuration has compact spatial sections (and,
consequently, no spin from isospin effect) it possesses interesting features
which are worth mentioning.\footnote{See \cite{IS} for the construction of
gravitating merons in $D$-dimensional massive Yang-Mills theory and the Skyrme
model.}

We consider a static space-time metric that is a product of $\mathbb{R}\times
S^{3}$ with a constant scale factor $\rho_{0}$, namely
\begin{equation}
ds^{2}=-dt^{2}+\frac{\rho_{0}^{2}}{4}\biggl((d\gamma+\cos\theta d\phi
)^{2}+d\theta^{2}+\sin^{2}\theta d\phi^{2}\biggl)\ , \label{S3}%
\end{equation}
together with the following ansatz for the gauge field
\begin{equation}
F_{1}(x^{\mu})=\gamma\ ,\qquad F_{2}(x^{\mu})=\theta\ ,\qquad F_{3}(x^{\mu
})=\phi\ , \label{Usol}%
\end{equation}
and
\begin{equation}
\lambda=\frac{1}{2}\ .
\end{equation}
With the above ansatz the $SU(N)$-YM equations are identically satisfied,
while the Einstein equations provide the following constraints between the
coupling constants
\begin{equation}
\rho_{0}^{2}=\frac{\kappa T_{\text{N}}}{e^{2}}\ ,\qquad\Lambda=\frac{3}%
{2}\frac{e^{2}}{\kappa T_{\text{N}}}\ .
\end{equation}
The energy density of the soliton is then
\begin{equation}
T_{00}=\frac{3}{2}\frac{T_{\text{N}}}{e^{2}\rho_{0}^{4}}=\frac{\Lambda}%
{\kappa}=\frac{3}{2}\frac{e^{2}}{\kappa^{2}T_{\text{N}}}\ .
\end{equation}
One can see that, if one requires having a static gravitating
configuration, then the cosmological constant must scale as $1/N$, so that it
must be small and positive when $N$ is large.

One can also consider a time-dependent scale factor, $\rho=\rho(t)$, in which
case the field equations read
\begin{gather}
\ddot{\rho}-\frac{1}{3}\Lambda\rho+\frac{\kappa T_{\text{N}}}{2e^{2}\rho^{3}%
}\ =\ 0\ ,\\
\dot{\rho}^{2}-\frac{1}{3}\Lambda\rho^{2}+1-\frac{\kappa T_{\text{N}}}%
{2e^{2}\rho^{2}}\ =\ 0\ .
\end{gather}
The above equations system represents a cosmological space-time whose source
is the energy-momentum tensor of a non-embedded $SU(N)$ meron, because still
in this dynamical case the YM equations are identically satisfied for
$\lambda=1/2$. We hope to come back on the analysis of these cosmological
space-time in a future publication.


\section{Conclusions and perspectives}


In this paper we have constructed meron BHs and self-gravitating soliton
solutions in the Einstein $SU(N)$-YM theory in $D=4$ and $D=5$ dimensions for
all values of $N$. These analytic configurations have been found by combining
the generalized hedgehog ansatz with the Euler parameterization of the $SU(N)$
group from which the YM equations are automatically satisfied for all values
of $N$, while the Einstein equations can be solved analytically.

One of the main results of this work is that we explicitly show the role that
the color number $N$ plays in the gravitational spin from isospin effect. In
fact, meron BHs can be distinguished by colored BHs by looking at the spin
from isospin effect, because this effect is present only in the meron BHs
constructed here.

In order to compute the spin generated from the isospin we have considered a
Bosonic field charged under $SU(N)$ around the background gauge field of the
BH solutions, showing that this mechanism works differently for the BHs in
$D=4$ and $D=5$. This difference lies in the presence of a non-zero
topological charge for the ansatz of the $D=5$ case.

Also, using the theory of non-embedded ansatz for $SU(N)$ together with the
spin from isospin effect, one can build fields of arbitrary high spin out of
scalar fields charged under the gauge group. Hence, one can analyze
interacting higher spin fields in asymptotically flat space-times without
introducing by hand higher spin fields.

\section*{Acknowledgments}

The authors would like to thank M. Valenzuela and F. Novaes for useful
discussions and suggestions. F. C. has been funded by Fondecyt Grant No.
1200022. M. L. is funded by FONDECYT post-doctoral Grant No. 3190873. A. V. is
funded by FONDECYT post-doctoral Grant No. 3200884. The Centro de Estudios
Cient\'{\i}ficos (CECs) is funded by the Chilean Government through the
Centers of Excellence Base Financing Program of ANID.

\end{document}